# Design of a compact low loss 2-way millimetre wave power divider for future communication

[1.]Muhammad Asfar Saeed, [1]Augustine O. Nwajana, [2]Muneeb Ahmad

[1]University of Greenwich, UK.

[2]Kumoh National Institute of Technology, South Korea.

## Abstract

In this paper, a rectangular-shaped power divider has been presented operating at 27.9 GHz. The power divider has achieved acceptable results for important parameters such as S11, S12, S21, and S22. The substrate employed for the power divider is Roger 3003 which has a thickness of 1.6 mm. This power divider provides a reflection coefficient of -12.2 dB and an insertion loss of 3.1 dB at 28 GHz. This ka-band T-junction power divider covers 68% of the bandwidth. Dimensions of the ka-band T-junction power divider are 50x80 mm. Due to its dimensions and bandwidth this power divider is more suitable for millimetre wave applications like RADAR, beamforming, and 5G applications.

1. Introduction

Wireless communication devices are increasing rapidly as the modern world is heading towards the autonomous era, which includes autonomous vehicles and IoT-based home appliances. All these devices need low-loss high-rate data connectivity for flawless operations. Current cellular wireless technology struggles to meet the growing demands of users due to the rapid increase in data consumption, driven by the rising number of consumers and the surge in average traffic per user. This led researchers to think of a next-generation cellular network known as 5G. While previous generations have predominantly relied on prime spectrum below 2 GHz, this frequency range is becoming increasingly congested. The millimetre-wave (mm-Wave) frequency has been proposed to address this limitation as a complementary solution to the already saturated lower bands [1]. The need for higher data rates and the utilization of these higher frequencies necessitates the implementation of beamforming techniques in both mobile devices and base stations within the cellular network. 5G needs five essential technologies to empower the fifth generation: small cells, full-duplex, beamforming, MIMO, and mm-wave [2]. The primary objectives of this emerging technology are to deliver high data rates to individual users and accommodate a substantial number of users within each cell. The most feasible strategy to increase network capacity is to utilize a broader spectrum. However, given the current congestion in the microwave band spectrum, the 28-GHz band emerges as a promising candidate for 5G deployment [2]. To fully utilize the 28-GHz band frequencies, it is essential to implement an antenna array integrated with a beamforming network. To achieve sidelobe compression, enhance array gain, and minimize interference, a power divider is required to precisely adjust the excitation amplitudes of the array's antenna elements.

In literature, several power dividers have recently been proposed for 28-GHz band applications [3]–[4]. Tuneable In-Phase Power Divider for 5G Cellular Networks proposed in [3] The designed power divider offers a wide range of frequencies. Similarly, a Tuneable millimetre-wave power divider for future 5G cellular networks is proposed in [5]. This paper introduces a modified Wilkinson power divider designed to function in the millimetre-wave band, specifically at 28 GHz, for future 5G mobile systems. In another technique a compact, wideband four-way power divider designed for operation in the W-band, utilizing a Riblet-type coupler [6]. The performance of this divider is compared against a conventional four-way power divider, which is constructed by connecting two two-way E-plane power dividers to the output of another two-way E-plane power divider, or by using a turnstile power divider [6]. The above-mentioned designs have achieved acceptable results but to fulfil the needs of the next generation. A 5G-supportive power divider for beamforming must exhibit low insertion loss to ensure efficient power distribution, high isolation to prevent interference between output ports, and excellent amplitude and phase balance for accurate signal combining. It should operate reliably across the wide 5G frequency range (e.g., 24-40 GHz) with minimal signal distortion. Additionally, the power divider must handle high power levels typical in 5G systems while maintaining low return loss for optimal impedance matching with a compact size.

In this paper, a power divider proposed fulfils the requirements of the next-generation network to support beam steering. Following the introduction, this paper is structured into three further sections. Section 2 covers the design and development of the proposed power divider, emphasizing its construction and key characteristics. Section 3 provides a thorough analysis of the power divider, examining parameters such as S11, S12, S21, S22, efficiency, and surface current. Finally, Section 4 summarizes the findings and offers conclusions based on the study's outcomes. Each section contributes to a comprehensive understanding of the power divider's design, performance, and relevance in modern communication systems

## 2. Design and development of power divider

Designing dividers for the Ka-frequency band (28GHz) using microstrip technology involves applying microwave engineering principles to ensure efficient power distribution. As illustrated in Fig. 1, microstrip technology has been selected for the power divider layout due to its compact size, ease of fabrication, and cost-effectiveness. The configuration of the proposed power divider is shown in Fig. 1(a). The design is developed on the substrate Roger 3003 with dielectric constant = 3.0 and thickness = 0.017 mm. The designed power divider covers an overall area of 50x80 mm.

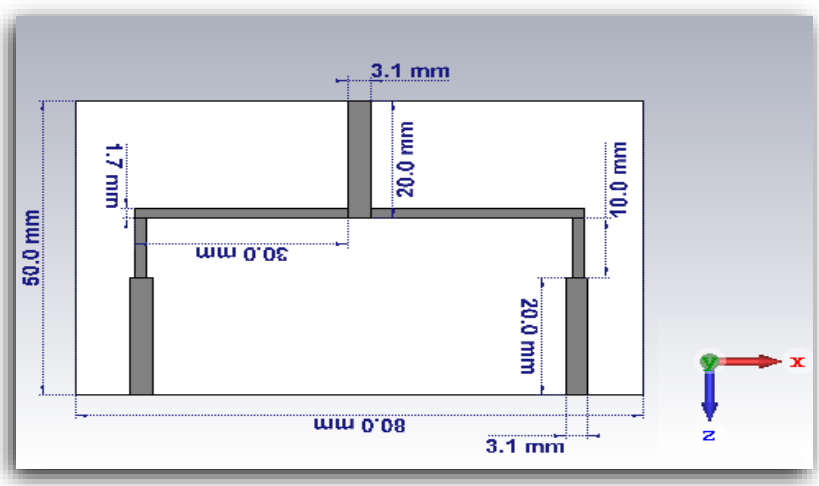

**Figure 1** Design of the Power Divider

The designed power divider is rectangular in shape, consisting of Roger 3003 as dielectric substrate material and utilizes copper as a conducting material. The top and bottom layers consist of copper whereas the Roger 3003 is placed between the copper layers in the middle. The characteristic impedance $Z_o$ of the transmission lines in a power divider is set to 50 ohms to match the system impedance. The quarter-wave transformers in the divider have a characteristic impedance calculated using:

$$Z_{(quarter-wave)} = Z_0 \sqrt{2} \tag{1}$$

For a typical 50-ohm system:

$$Z_{(quarter-wave)} = 50 \times 2 \approx 70.7 \, \Omega \tag{2}$$

The length of each transmission line section is designed to be a quarter-wavelength (λ/4) at the operating frequency of 28 GHz. The wavelength (λ) is determined by:

$$V = f\lambda \tag{3}$$

where:

c is the speed of light ($3 \times 10^8$ m/s), f is the frequency (28 GHz).

Thus

$$\lambda = \frac{3 \times 10^8}{28 \times 10^9} \approx 0.017 m \tag{4}$$

The quarter wavelength is:

$$\frac{\lambda}{4} \approx 0.002675 m \, (2.675 mm) \tag{5}$$

The designed power divider has two 70Ω transformers, both transformers are connected to 50Ω connectors. The designed power dimensions are explicit in Table 1, whereas the Figure shows the simulated power divider.

**Table I** Dimensions of the designed Power Divider

| Parameters | Symbols | Dimensions (mm) |
|---|---|---|
| **Input transmission line** | $TL_{(L)}$ | 20 |
| | $TL_{(w)}$ | 3.1 |
| | $TL_{(H)}$ | 0.017 |
| **1st Quarter wave transformer (70Ω)** | $QWT1_{(L)}$ | 1.7 |
| | $QWT1_{(W)}$ | 30 |
| **2nd Quarter wave transformer (70Ω)** | $QWT2_{(L)}$ | 1.7 |
| | $QWT2_{(W)}$ | 30 |
| **1st connector (50Ω)** | $C1_{(L)}$ | 20 |
| | $C2_{(w)}$ | 3.1 |
| **2nd connector (50Ω)** | $C1_{(L)}$ | 20 |
| | $C2_{(W)}$ | 3.1 |
| **Substrate and Ground** | L | 50 |
| | W | 80 |
| | H | 1.6 |

## 3. Results and analysis:

The design power divider is analysed using important parameters like Reference impedance, reflection coefficient, insertion loss, isolation loss, and surface current.

### a) S11 (Reflection Coefficient):

S11 represents the reflection coefficient at the input port. It is the ratio of the reflected power to the incident power at the input port. S11 indicates how much of the signal is reflected towards the source at Port 1. It is used to assess the impedance matching of the device at the input port. A low S11 value (close to 0 dB) indicates good matching and minimal reflection, while a high S11 value indicates poor matching and high reflection. The designed power divider has achieved a -12 dB reflection coefficient. The simulated result is shown in the Figure.

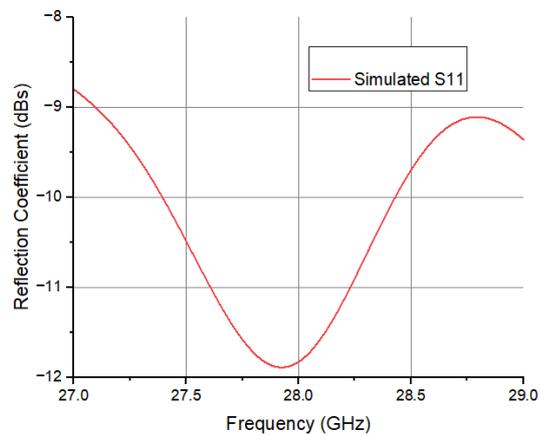

**Figure 2** Reflection Coefficient of power divider

### b) S12 (Reverse Transmission Coefficient):

S12 represents the reverse transmission coefficient. It is the ratio of the power transmitted from Port 2 back to Port 1. S12 indicates how much of the signal applied at Port 2 is transferred to Port 1. In an ideal one-way device (like an amplifier), S12 should be very low (indicating little to no signal is transferred from output back to input). The designed power divider S12 simulated result is illustrated in Figure.

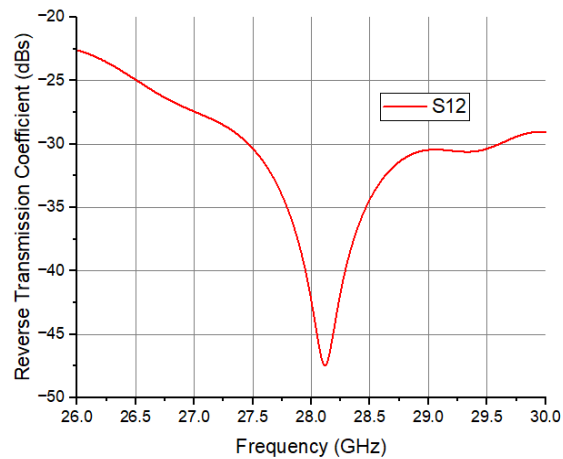

**Figure 3** Reverse Transmission Coefficient of power divider

c) **S21 (Forward Transmission Coefficient)**

S21 represents the forward transmission coefficient. It is the ratio of the power transmitted from Port 1 to Port 2. S21 indicates how much of the signal applied at Port 1 is transferred to Port 2. It is a key parameter in determining the gain or loss of the network. A higher S21 value indicates better transmission from input to output.

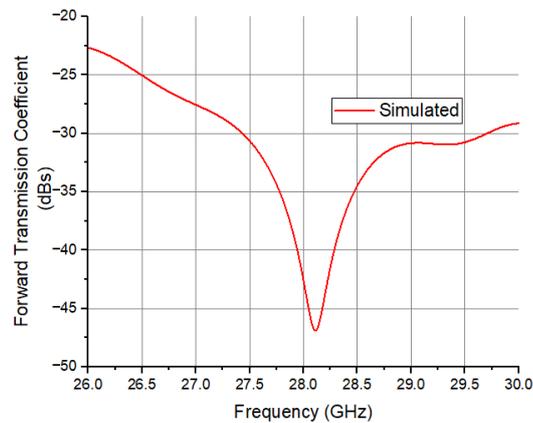

**Figure 4** Forward Transmission Coefficient of power divider

d) **S22 (Output Port Reflection Coefficient)**

S22 represents the reflection coefficient at the output port. It is the ratio of the reflected power to the incident power at the output port (Port 2). S22 indicates how much of the signal is reflected back towards the source at Port 2. Similar to S11, it is used to assess the impedance matching of the device at the output port. A low S22 value indicates good matching and minimal reflection, while a high S22 value indicates poor matching and high reflection. The figure shows the S22 simulated results.

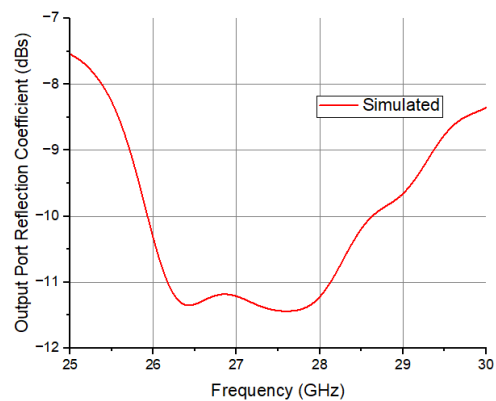

**Figure 5** Output port Reflection Coefficient of power divider

e) **Surface current:**

It is evident from the figure that when an RF signal is applied to the input port, the surface current begins to flow through the input transmission line toward the T-junction where the signal splits. The current at the input port is evenly distributed, indicating that the divider is ready to split the power equally between the two output ports. In the T-junction, the input current is divided into two equal parts, each traveling down one of the two branches leading to the output ports. The surface current is nearly identical on both branches, indicating a balanced power split as shown in the figure. It is also noticeable from the simulated results that surface current on each branch is strong and consistent, with minimal variations, suggesting efficient power division and low signal loss. Each branch of the power divider includes a quarter-wave transformer to match the impedance between the input and the output ports. The surface current along these transformers is uniform, without sharp drops or spikes, which indicates effective impedance transformation and minimal reflection. Any irregularities in the surface current in these areas might indicate potential losses or mismatches.

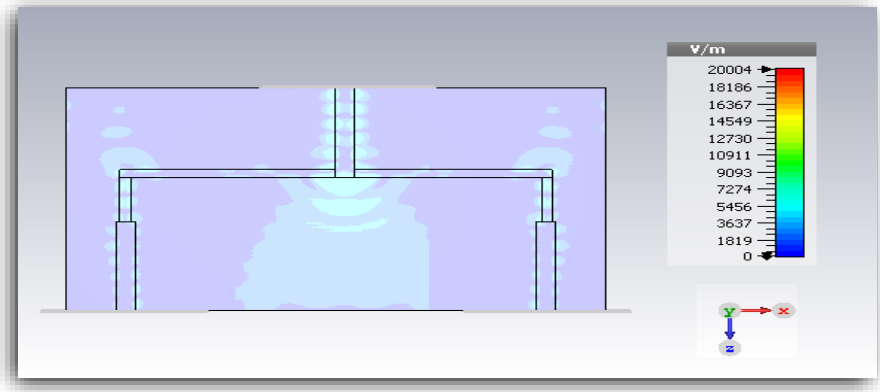

**Figure 6** Surface Current of the power divider

**Table II** Comparison table of the power divider

|  | Number of ports | Frequency | Fractional bandwidth | Insertion loss | Size (mm) | Material | Structure | Reflection coefficient |
|---|---|---|---|---|---|---|---|---|
| [12] | 4 | 2.5 | 80% | 3 | 40x50 | Roger5880 | Y-Junction | -13 |
| [13] | 3 | 6.8 | 61% | 0.4 | 80x110 | FR-4 | T-Junction | -12 |
| [14] | 3 | 18 | 47% | 1 | 50x50 | ceramic-filled PTFE | Circular junction | -13 |
| [15] | 4 | 26 | 52% | 0.7 | 112x150 | FR-4 | T-Junction | -15 |
| [16] | 5 | 7-12 | 64% | 0.5 | 75x80 | RT Duroid | T-Junction | -14 |
| [17] | 4 | 28 | 54% | 3.7 | 30x50 | RT Duroid | Circular | -12 |
| Proposed work | 3 | 27.9 | 68% | 3.1 | 50x80 | Roger 3003 | T-junction | -12 |

## 4. Conclusion:

In this paper, a microstrip power divider has been designed to operate at 27.9GHz frequency. The designed power divider is compact and rectangular-shaped. Roger 3003 and copper are used for the modelling of the power divider. The power divider can split the input power into two outputs equally which is 50Ω per connector. The designed power divider consists of two transformers and two connectors, overall, it covers an area of 50x80mm. The simulated results of the designed microstrip power divider illustrate perfect matching and offer acceptable results of important parameters like S11, S12, S21, and S22.